# *Property unification of inherent amplitude, phase and polarization within a light beam*

Xiaoyu Weng[1]*, Yu Miao[2], Guanxue Wang[2], Yihui Wang[3], Qiufang Zhan[2], Xiangmei Dong[2], Junle Qu[1], Xiumin Gao[2]* and Songlin Zhuang[2]

*1 Key Laboratory of Optoelectronic Devices and Systems of Ministry of Education and Guangdong Province, College of Physics and Optoelectronic Engineering, Shenzhen University, Shenzhen, 518060, China.*

*2 Engineering Research Center of Optical Instrument and System, Ministry of Education, Shanghai Key Lab of Modern Optical System, School of Optical-Electrical and Computer Engineering, University of Shanghai for Science and Technology, 516 Jungong Road, Shanghai 200093, China.*

*3 Lee Shau Kee School of Business，Hong Kong Metropolitan University，Hong Kong，999077, China.*

* Correspondence and requests for materials should be addressed to X.W. (email: xiaoyu@szu.edu.cn ) or to X.G. (email: gxm@usst.edu.cn ).

# Abstract:

Is it possible to modulate the inherent properties of a single light beam, namely amplitude, phase and polarization, simultaneously, by merely its phase? Here, we solve this scientific problem by unifying all these three properties of a single light beam using phase vectorization and phase version of Malus's law. Full-property spatial light modulator is therefore developed based on the unification of these fundament links, which enables pixel-level polarization, amplitude and phase manipulation of light beams in a real-time dynamic way. This work not only implies that the amplitude, phase and polarization of a single light beam are interconnected, but also offers a solid answer on how to modulate these three natures of a single light beam simultaneously, which will deepen our understanding about the behavior of light beam, and facilitating extensive developments in optics and relate fields.

# 1 Introduction

Is it possible to modulate the inherent properties of a single light beam, namely amplitude, phase and polarization, simultaneously, by merely its phase? The key to solve this scientific problem relies on the unification of these three attributes of a single light beam. As we know, amplitude, phase and polarization are three properties of light beam. The former two are the scalar properties, while the latter refers to the electric oscillation, and is a vector property. Throughout the development of optics, the relationship between these three natures are the basis of manipulating light beam. Therefore, in the past decades, every breakthrough regarding the establishment of fundamental links between amplitude, phase and polarization has not only deepened our understanding about the behavior of light beam, but also given rise to wide-range of applications, including optical communications [1, 2], advanced lasers [3-5], optical storage [6], optical displays [7, 8], and optical imaging [9-11].

Generally, there are three familiar relationships between amplitude, phase and polarization in classical optics, namely, the mutual link between phase and amplitude, the polarization-to-amplitude link and the polarization-to-phase link. Specifically, the mutual link between phase and amplitude can be established by the wavefront technique [12-16], which is the cone stone of scalar optics; the polarization-to-amplitude link demonstrated by the renowned Malus's law implies that modulating the polarization of a light beam can be turned into amplitude adjustment with the aid of a polarizer; the Pancharatnam-Berry (PB) phase in 1987 demonstrates the polarization-to-phase link [17]. Based on this principle, one can convert a right-hand circularly polarized beam into a left-hand circularly polarized one along with an additional phase and vice versa. These above three fundamental links imply that vector polarization can link to the scalar phase and amplitude, but not the opposite. In our previous work, we solve this scientific problem using the principle of phase vectorization [18]. Accordingly, the phase-to-polarization link is established in classical optics, and the scalar phase of a single light beam satisfied the wave function can link with its vector polarization directly. Even so, we can only assert that only two of amplitude, phase and polarization are interconnected. That is, these three properties of the light beam are relatively independent, and all the above four fundamental links are, to some extent, viewed as four independent links. Can we connect these four fundamental link so that all these three natures, namely phase, amplitude and polarization, of a single light beam unify as a whole?

Here, we demonstrate the unification of amplitude, phase and polarization of light beam using the principle of phase vectorization and phase version of Malus's law. Taking $m$ order vector vortex beam (VVB) as example. Based on these two principles, three kinds of phases, named Phase A, B, C, are linked to the amplitude, phase and polarization of VVB, respectively. That is, all these three natures of VVB can be controlled merely by its scalar phase. Full-property spatial light modulator (SLM) is therefore developed based on the unification of amplitude, phase and polarization, which enables pixel-level polarization, amplitude and phase manipulation of light beams in a real-time dynamic way. The work reports the invention of Full-property SLM that offers a solid answer on how to modulate the phase, amplitude and polarization of a single light beam simultaneously using only its phase property. More importantly, this property unification of light beam implies that amplitude, phase and polarization of light beam satisfied the wave function are interconnected.

# 2 Results

## 2.1 Main scientific goal

Although there are four fundament relationships between the inherent amplitude, phase and polarization of a light beam, these three natures of light beam are considered to be independent from each other because only every two properties are interconnected [18]. Here, our main scientific goal is to unify these three natures using merely the scalar phase of light beam, as shown in Fig. 1. It should be emphasized that three properties of light beam can only link with three kinds of phases. Therefore, we should find out three kinds of phases that are responsible for adjusting amplitude, phase and polarization of light beam, respectively. More importantly, this property unification requires that these three kinds of phase are not affect each other. Otherwise, full property modulation of light beam cannot be achieved. For simplicity, we name these three kinds of phases as Phase A, B and C, which link to the polarization, amplitude and phase of light beam, respectively.

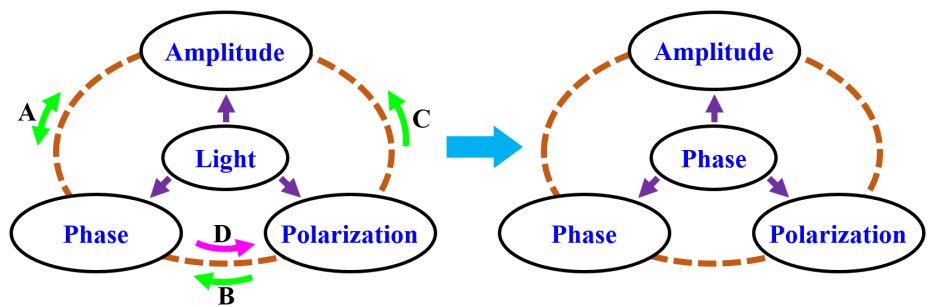

**Figure 1.** Main scientific goal of our work. Here, there are four fundamental relationships between amplitude, phase and polarization of light: arrow A denotes the mutual relationship between amplitude and phase; arrow B indicates the polarization-to-phase link; and arrow C denotes the polarization-to-amplitude link. Arrow D represents the phase-to-polarization link [18]. Our main scientific goal is to unify amplitude, phase and polarization of a single light beam using merely its scalar phase.

## 2.2 Phase A-to-polarization link

As already demonstrated in our previous work, phase vectorization is capable of vectorizing the scalar phase of light beam into its vector polarization, thereby establishing the fundamental relation between phase and polarization [18]. It should be emphasized that the phase and polarization are both two inherent properties of a light beam. To realize this phase-to-polarization link, two critical conditions must be satisfied: one is that the light beam must possess an inherently different polarization response from the left and right circularly polarization modes; another is that the undesired circularly polarization mode must be eliminated without affecting the desired one. The first condition implies that not all of light beams are suitable for the establishment of phase-to-polarization link. Because the phase is a scalar property, while the polarization is a vector property. The second condition indicates how to extract the desired polarization mode from a single light beam.

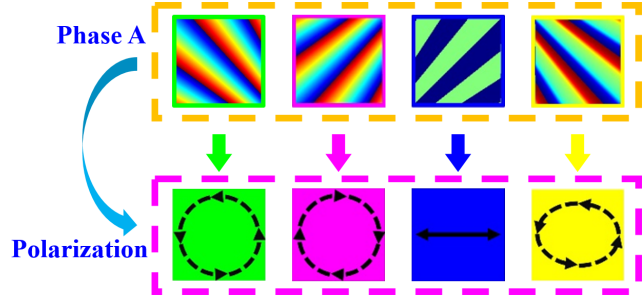

**Figure 2.** Physical essence of phase vectorization. Phase vectorization presents a one-to-one correspondence between the scalar phase and vector polarization of a single light beam, thereby transferring four specifics phases, namely vortex phases, binary phase and a combination of both above, into left and right circular, linear, and elliptical polarization, respectively. Note that the arrow indicates the polarization direction.

According to both conditions, we take *m*-order VVB as example to vectorize its scalar phase into its vector polarization. Therefore, one can simply obtain a one-to-one correspondence between the phase and polarization of *m*-order VVB. That is, the phase in Eq. (1) can be vectorized into the vector polarization in Eq. (2) [18].

$$\phi_A = \text{Phase}[\cos\varphi_0 \exp[i(m\varphi - \beta)] + \sin\varphi_0 \exp[-i(m\varphi - \beta)]]. \tag{1}$$

$$\mathbf{E} = \cos\varphi_0 \exp(-i\beta)|\mathbf{L}\rangle + \sin\varphi_0 \exp(i\beta)|\mathbf{R}\rangle. \tag{2}$$

Here, the parameter $\varphi_0$ in Eq. (1) is a weight factor that adjusts the proportion of left and right circular polarization mode $|\mathbf{L}\rangle$ and $|\mathbf{R}\rangle$. As shown in Fig. 1, the physical essence of phase vectorization is to transfer the polarization change into four specific phases: binary phase $\phi = \text{Phase}[\cos(m\varphi - \beta)]$ for linear polarization, vortex phases $\phi = \pm(m\varphi - \beta)$ for left and right circular polarization, and a combination of both for elliptical polarization. Therefore, we call the phase in Eq. (1) as Phase A, which is responsible for the polarization modulation of light beam.

## 2.3 Phase B-to-amplitude link

The renowned Malus's law demonstrates that modulating the polarization of a light beam can be turned into amplitude adjustment with the aid of a polarizer, thereby establishing the polarization-to-amplitude link. As shown in Supplementary Fig. 1, after passing through a polarizer, the light intensity of incident linearly polarized light beam turns into $I_o = |\mathbf{E}_0|^2 \cos^2\omega$. That is, the loss of light intensity can be obtained by $I_e = 1 - I_o = |\mathbf{E}_0|^2 \sin^2\omega$. Here, $\mathbf{E}_0$ is the amplitude of incident light beam and $\omega$ is the angle between the polarization direction of incident light beam and polarizer. As the polarizer rotates, both parts of light intensities $I_o$ and $I_e$ transform each other by $\omega$. Therefore, it is reasonable to consider that $I_o$ and $I_e$ are two complementary intensity modes within a light beam that are intertwined with each other during propagating in free space. The polarizer is utilized to separate them from each other and extract $I_o$. In this way, one can obtain a one-to-one correspondence relationship between polarization and amplitude of light beam.

The Malus's law implies an important physical idea that involves how to extract the inherent intensity mode $I_o$ from a light beam. Thus, to extend this polarization-to-amplitude link into its phase version, the inherent intensity mode $I_o$ must be extracted using the phase of light beam. As discussed above, $I_o$ and $I_e$ can be considered as two complementary intensity modes. During propagating in free space, both scalar modes are intertwined with each other so that the entire light beam remains stable. Although $I_o$ cannot be extracted directly, we still can achieve an indirect scalar mode extraction of a light beam. Specifically, based on the sample optical system of phase vectorization in Fig. 5, $I_o$ and $I_e$ can be separated spatially in the focal region of OL₁ using optical pen [16], where $\omega$ can be adjusted by the phase of incident light beam. After passing through a pinhole, $I_e$ is eliminated, and only $I_o$ is retained. In this way, the phase-to-polarization link is established, thereby linking the phase B in Eq. (3) to the amplitude in Eq. (4). Detail derivation can be found in Supplementary Note 1.

$$\phi_B = \text{Phase}\left(\sum_{j=1}^{N}\left(Amp_{aj} + Amp_{bj}\right)\times \text{PF}(1, f_j, \eta_j, 0, 0)\right) \tag{3}$$

$$I_o = \mid E_0 \mid^2 \cos^2 \omega \tag{4}$$

where

$$Amp_{aj} = \text{PF}(1, -f_j, \eta_j, 0, \omega) + \text{PF}(1, f_j, \eta_j, 0, -\omega + 0.5\pi) \; ; \tag{5}$$

$$Amp_b = \text{PF}(1, -f_j, \eta_j, 0, -\omega) + \text{PF}(1, f_j, \eta_j, 0, \omega + 1.5\pi) \; . \tag{6}$$

Here, $N$ indicates the number of complementary intensity modes pairs, namely $I_o$ and $I_e$, in the focal region of OL₁. $\text{PF}(s_j, f_j, \eta_j, z_j, \delta_j)$ denotes the optical pen [16]. $\left(f_j, \eta_j, z_j = 0\right)$ indicate the position of $j$-th focus in the cylindrical coordinate system; $s_j=1$ and $\delta_j$ are weight factors that can be used to adjust the amplitude and phase of the $j$-th focus, respectively.

By comparing with Malus's law, the phase B-to-amplitude link indicates an indirect amplitude mode extraction $I_o$ of a light beam. The pinhole in Fig. 5 acts as a polarizer of Malus's law, which is responsible to eliminate the undesired amplitude mode $I_e$. For this reason, we call this fundamental link as phase version of Malus's law, namely phase Malus's law. As shown in Fig. 3, Phase Malus's law converts a series of phase into the amplitude of light beam. Every particular value of amplitude is correspondent with one particular kind of phase, the distribution of which is determined by the parameter *N*. Based on the phase Malus's law, the amplitude change of light beam is no longer dependent on the polarization direction of polarizer, but the phase of incident light beam. Therefore, one can simply achieve pixel amplitude modulation of light beam using pixel phase modulation device, such as phase-only SLM.

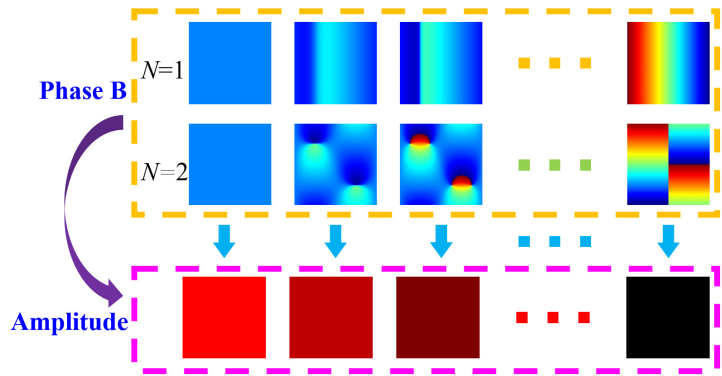

**Figure 3.** Physical essence of phase Malus's law. The phase version of Malus's law, namely phase Malus's law, implies a one-to-one correspondence between the phase and amplitude of a single light beam. Therefore, every particular value of amplitude is correspondent with one particular kind of phase, the distribution of which is determined by the parameter *N*.

## 2.4 Unification of amplitude, phase and polarization

So far, we have established the phase A-to-polarization link and the phase B-to-amplitude link by the principle of phase vectorization and phase Malus's law. In principle, both fundamental links are all realized based on one physical idea, namely inherent mode extraction of a light beam. Phase vectorization implies a vector polarization mode extraction, while phase Malus's law denotes a scalar amplitude mode extraction. Unlike the phase vectorization, phase Malus's law is valid for every light beam in physics, including the vector light beam using for phase vectorization. That is, Phase A and B do not affect each other because they represent different properties of light beam, namely the polarization and amplitude, respectively. As shown in Fig. 4, we therefore unify the amplitude, phase and polarization of a light beam by merely its scalar phase, which can be expressed as

$$\phi = \phi_A + \phi_B + \phi_C .\tag{7}$$

Here, Phase C indicates other phases, which is responsible for the pure phase modulation. It should be emphasized that the light beam mentioned in this paper is the solution of wave function.

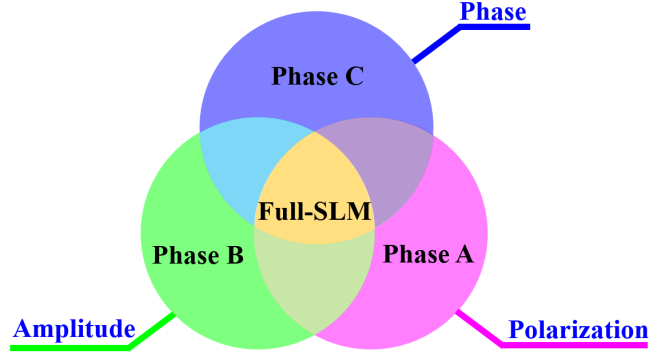

**Figure 4.** Unification of amplitude, phase and polarization. Based on the property unification of light beam, three kinds of phases, namely Phase A, B, and C, are corresponding with the polarization, amplitude and phase of light beam, respectively.

## 2.5 Full-property SLM

Based on the unification of amplitude, phase and polarization in Fig. 4, we establish a Full-property SLM. Full-property SLM shares a same optical system with polarized-SLM, which can also be simplified to a filter system in Fig. 5 [18]. In the entire optical system of Full-property SLM, a collimated incident x linearly polarized beam with a wavelength of 633 nm propagating along the optical axis is converted into a $m$=30 order VVB by a vortex polarizer (VP) after modulating by a phase-only SLM. The VP can be easily manufactured using the Q-plate technique [19, 20]. When focusing by the first objective lens OL$_1$, the modulated VVB coded by the phase A, B and C are divided into desired and undesired modes, respectively. Generally, phase A is corresponding to the desired vector polarization modes $\exp(-i\beta)|\mathbf{L}\rangle$, $\exp(i\beta)|\mathbf{R}\rangle$ and the undesired vector polarization modes $\exp[i(2m\varphi - \beta)]|\mathbf{R}\rangle$, $\exp[-i(2m\varphi - \beta)]|\mathbf{L}\rangle$, respectively. Phase B is related to the desired scalar amplitude mode $I_o$ and the undesired scalar amplitude mode $I_e$, respectively. Because all desired vector and scalar modes are in the geometric focal position of OL$_1$, while their counterpart undesired ones are located at the position far away from the desire modes, one can simply obtain the desired modes using a pinhole with a radius of 400 μm. That is, the desired polarized modes in the focal region of OL$_1$ have one-to-one correspondence with the phase A, B and C. When reconstructed by the second objective lens OL$_2$, the amplitude, phase and polarization output from Full-property SLM link with the VVB phase directly as well. In this way, the amplitude, polarization and phase of VVB are unified by merely its scalar phase. Here, the numerical apertures (NA) of both OL$_1$ and OL$_2$ are 0.01.

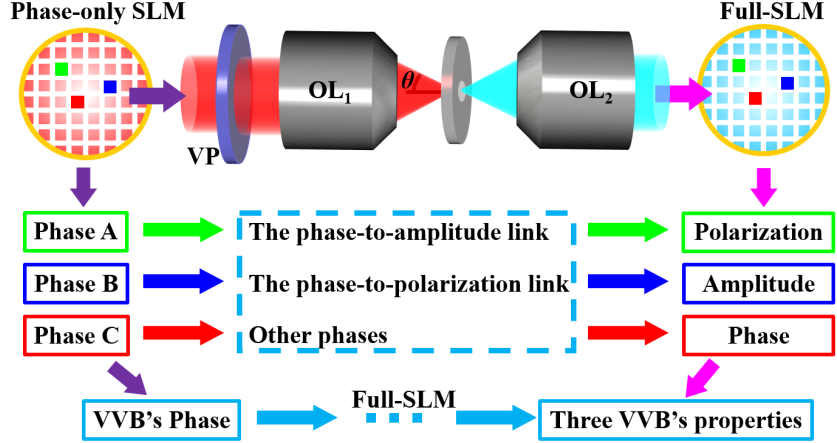

**Figure 5.** Schematic of full-property SLM. A collimated incident x linearly polarized beam is converted into a *m*=30 order VVB by a vortex polarizer (VP) after modulating by a phase-only SLM. Based on the property unification, the amplitude, polarization and phase of light beam can be adjusted simultaneously using full-property SLM.

## 2.6 Pixel-level modulation using Full-property SLM

Based on the above property unification of light beam, one can simply adjust its three natures merely by its scalar phase. Generally, Phase A, B and C can be pixelized by the phase-only SLM in Fig. 5. The pixelate phase enables pixelate polarization, thereby permitting individual adjustment of all these three properties in each pixel. In Fig. 6, we present three theoretical examples of performing pixel-level full-property modulation of *m*=30 order VVB using Full-property SLM. The wavefront distribution of *m*=30 order VVB is shown in Fig. 6 (a). In the following simulations, we demonstrate how to modulate the amplitude, phase and polarization of VVB in Fig. 6 (a) merely by its scalar phase. Detail theoretical principle is presented Supplementary Note 2.

### 2.6.1 Amplitude modulation of incident *m* order VVB

Figure 6 (b-d) present a theoretical result of amplitude modulation of incident *m* order VVB. The VVB phases possess two kinds of zones, namely zones A, B and zones C, D, E, F. The first kind, namely zones A, B, are x linearly polarized, while the second kind, namely zones C, D, E, F, are y linearly polarized. Both polarizations are easily realized by the phase A, which is responsible for polarization modulation in Eq. (1). After passing through a polarizer indicated by the yellow arrow, only one kind of zones are retained, as shown in Fig. 6 (c, d). In term of amplitude modulation, each zone possesses different amplitude, which can be manipulated by phase B. Specially, the amplitude of zones A and B, can be expressed as

$$I_g = |\mathbf{E}_0|^2 \cos^2 \phi_{ZAB} \ , \tag{8}$$

while that of zones C, D, E, and F are

$$I_h = |\mathbf{E}_0|^2 \cos^2 \phi_{ZCDEF} \ , \tag{9}$$

Here,

$$\phi_{ZAB} = \begin{cases} 0; & zoneA \quad Phase[\cos(\varphi + k\sin\theta / NA)] = -1 \\ 0.5\pi; & zoneB \quad Phase[\cos(\varphi + k\sin\theta / NA)] = 1 \end{cases} ;$$

$$\phi_n = \cos n\varphi \ ;$$

$n$=1, 2, 3, and 4 represent the circular zones C, D, E, and F, respectively.

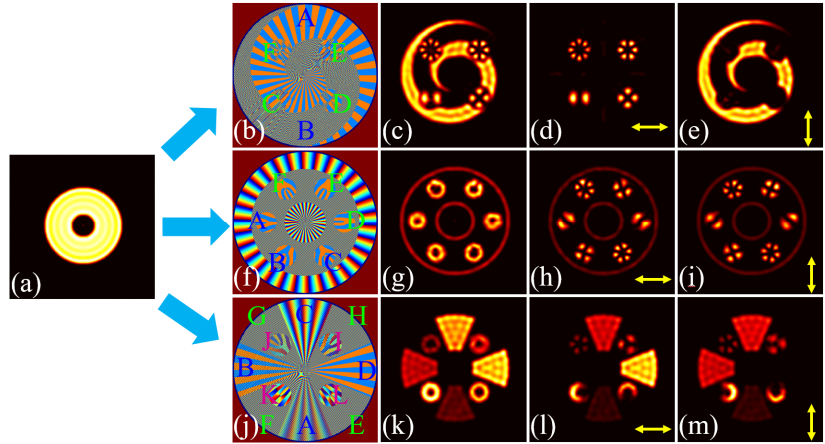

**Figure 6.** Theoretical result of pixel-level full-property modulation using Full-property SLM. Here, (a) wavefront light intensity of m order VVB. After modulating by the phases (b, f, j), one can obtain three examples of pixelate modulation of light beam: (c) amplitude modulation; (g) polarization modulation; (j) amplitude, polarization and phase modulation. (d, e), (h, i) and (l, m) are their corresponding light intensities distribution after passing through a polarizer indicated by the yellow arrow.

### 2.6.2 Polarization modulation of incident *m* order VVB

Figures 6 (b-d) mainly verify the amplitude modulation of Full-property SLM. By comparison, we provide a theoretical result of polarization modulation in Figs. 6 (f-i). Based on the full property unification of m order VVB, the Full-property SLM can transform *m* order VVB in Fig. 6(a) into the vector beam in Fig. 6 (g) by coding the phase in Fig. 6(f). In the transverse plane of light beam in Fig. 6 (g), there are eight different polarization states. The edges of entire light beam in Fig. 6 (a) indicated by inner and outer ring are left and right circularly polarized, respectively, which correspond to Phase A with $\phi = \pm(m\varphi - \beta)$ [see Eq. (1)], respectively. As shown in Fig. 6 (g), the amplitude within the light beam are zero except 6 circular zones. These zones are also named as zone A, B, C, D, E, and F, respectively. Light beams in these zones possesses the polarization state of VVB with different order, which are correspondent with the phase A in Fig. 6 (f). Here, the polarization of zones A, B, C can be expressed as

$$P_n = \begin{bmatrix} \cos n\varphi + \beta \\ \sin n\varphi + \beta \end{bmatrix}. \tag{8}$$

where $\beta = 0.25\pi$ ; $n$=1, 2, 3 represent zones A, B, C, respectively, and $n$=-1, -2, -3 represent zones D, E, F, respectively. Because of the plus-minus sign of n, after passing through the polarizer indicated by the yellow arrows, the petals with $2n$ numbers are rotated along with the polarizer reversely, as shown in Figs. 6 (h, i).

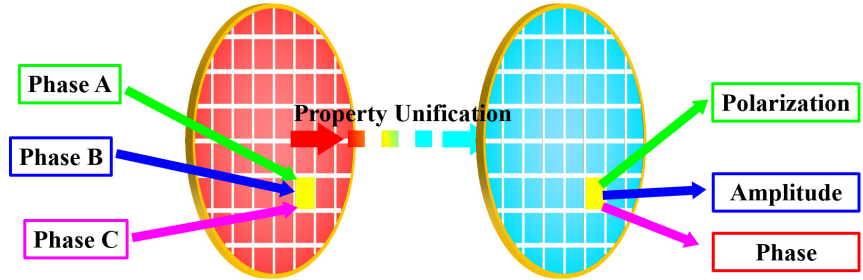

**Figure 7.** Schematic of full property modulation. Based on the property unification of light beam, the amplitude, polarization and phase of light beam can be modulated simultaneously in an identical pixel using merely its scalar phase, namely Phase A, B and C.

### 2.6.3 Full property modulation of incident $m$ order VVB

To some extent, Figures 6 (b, g) demonstrate the amplitude and polarization modulation of Full-property SLM. Theoretically, it is not a convincing answer for the scientific question: how to modulate the three natures of a single light beam, namely phase, amplitude and polarization, merely by its scalar phase. The reason is that phase A and B are spatially divided in the transverse plane of m order VVB. That is, one can only achieve those properties manipulation in different zone instead of an identical zone. It should be emphasized that full-property modulation of light beam requires to manipulate all these three natures, namely amplitude, phase and polarization, of light beam in an identical pixel by its scalar phase, as shown in Fig. 7. Therefore, we verify the full-property modulation of light beam using Full-property SLM in Figs. 6 (j-m). As shown in Fig. 6 (j), the phase of m order VVB possesses eight fatal-like zones, namely zones (A-D) and (E-H). The light intensities of zones (E-H) are adjusted to be zero, except the four small circular zones, namely zones (I-L). As shown in Fig. 6 (k), one can not only adjust the light intensity of one zone, but also modulate the polarization of an identical zone by overlapping Phase A with Phase B. Specifically, as shown in Figs. 6 (l, m), the polarization states of zones (A-D) are x linear polarization, left circular polarization, y linear polarization and right circular polarization, respectively, while their corresponding light intensities are 0.25, 0.5, 0.75, 1, respectively.

As discussed above, the polarization, amplitude, and phase are corresponding to the Phase A, B and C, respectively. When superposing all these three kinds of phases, one can obtain different light beam with different polarization, different phase and even different amplitude in the same zone of light beam, thereby providing a full property modulation in the zones (I-L), see Fig. 6 (k). The light beams in zones (I-L) are propagable vortex beams carrying natural non-integer orbital angular momentum. Their peculiar polarizations can be expressed as

$$\mathbf{E}_{mf} = \exp[i(l+0.5)\varphi]\begin{bmatrix} \cos[(m+0.5)\varphi + \beta] \\ \sin[(m+0.5)\varphi + \beta] \end{bmatrix}, \tag{9}$$

where $l$=-1, $m$=-2, $\beta = 0.5\pi$ for zone I; $l$=0, $m$=-3, $\beta = 0.5\pi$ for zone J; $l$=1, $m$=0, $\beta = 0.5\pi$ for zone K; $l$=2, $m$=-1, $\beta = 0.25\pi$ for zone L. According to Eq. (9), these vector beams not only possess complicate polarization state but also carry a natural optical vortex $\exp[i(l+0.5)\varphi]$. After passing through the polarizers indicated by the yellow arrow in Figs 6 (l, m), petals with number $2m$+1 are rotated along with the polarizer, which further demonstrates the polarization in each zone. Moreover, the amplitudes of zones (I, J, L, K) in Figs 6 (k) are adjusted to 0.25, 0.5, 0.75, 1, respectively. Note that the polarization distribution in zones (I-L) can be found in our previous work [21].

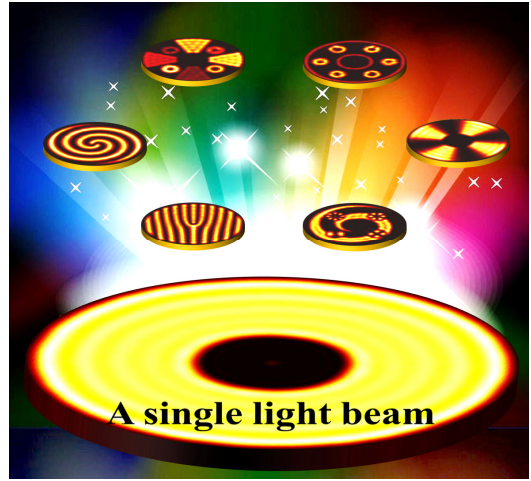

**Figure 8.** Schematic of inherent mode extraction of a single light beam. Both phase vectorization and phase version of Malus's law rely on one physical idea, namely extraction inherent arbitrary modes from a single light beam, including vector and scalar mode of light beam.

## 3 Discussion

### 3.1 Physical idea of full property unification of light beam

In the following, we would like to discuss the physical idea of full property unification of a single light beam. In our works, we are aim to unify the three natures, namely phase, amplitude and polarization, of a single light beam. Therefore, the phase-to-polarization link and the phase-to-amplitude link are established in classical optics by proposing the principle of phase vectorization and phase Malus's law. Frankly speaking, there is one physical idea that is the basis of both fundamental links, namely extraction inherent vector and scalar modes from a single light beam.

### 3.1.1 Vector mode extraction of a single light beam

Generally, in our optical text book, we are told that every new light beam can normally be created by superposing two orthogonal polarized beams with different phase in an interferometric optical system, which is also indicated by Jones matric theory. Based on this physical idea, one can only obtain one particular new light beam when the parameters of two orthogonal polarized beams are determined. That is, this conventional principle implies a one-to-one correspondence in the framework of generating a new light beam. No doubt, mathematically, one can always obtain 1+1=2, instead of other values. However, extraction inherent modes from a single light beam manifests an entirely inverse process of the above generalized principle. Similar to the mathematical form, we can not only obtain 2=1+1, but also 2=3-1 and 2=5-3 as well. Therefore, this inverse process provides plenty of possibilities to extract a desired light beam. This physical thought conveys an entirely new principle of polarization modulation, thereby making it possible to unify the amplitude, phase and polarization of light beam merely by its scalar phase.

### 3.1.2 Scalar mode extraction of a single light beam

Similarly, the conventional Malus's law implies a physical idea that involve how to extract the inherent intensity modes from a single light beam. Specifically, one can obtain $I_o = |\mathbf{E}_0|^2 \cos^2 \omega$ of a light beam by $I_e = |\mathbf{E}_0|^2 \sin^2 \omega$ using a polarizer. Both scalar modes $I_o$ and $I_e$ are two complementary intensity modes that are intertwined with each other during propagating in free space. The polarizer is utilized to separate them from each other and extract $I_o$. Derived from the identical physical idea of the conventional Malus's law, the Phase B-to-Amplitude link is a direct process of scalar mode extraction, but an indirect one using merely the phase of a single light beam, see Supplementary Note 1. For this reason, we call it the phase version of Malus's law, namely phase Malus's law. Although there are plenty of previous works that can also modulate the amplitude of light beam with its phase [12, 13], the phase version of Malus's law conveys a different physical idea, namely mode extraction of a light beam. Note that we are not concentrated on the technical details of above techniques, but their different physical idea.

### 3.2 Conclusion

In conclusion, we have theoretically demonstrated that all the three natures of light beam, namely amplitude, phase and polarization, can be unified by its scalar phase based on the principle of phase vectorization and phase Malus's law. Taking a *m* order VVB as an example. A Full-property spatial light modulator (SLM) is developed based on the full property unification of m order VVB, which enables pixel-level polarization, amplitude and phase manipulation of light beams in a real-time dynamic way. This work not only present a new principle of modulating light beam, namely the inverse process of the above conventional principle in an interferometric optical system, but also offers a solid answer on how to modulate the phase, amplitude and polarization of a single light beam simultaneously merely by its scalar phase. We believe the unification of the three natures of light beam will deepen our understanding about the behavior of light beam. That is, the properties of light beam are connected.

## Data Availability

All data supporting the findings of this study are available from the corresponding author on request.

**Acknowledgments**

Parts of this work were supported by the National Natural Science Foundation of China (62022059/11804232) and the National Key Research and Development Program of China (2018YFC1313803).

**Author contributions**

X. Weng conceived of the research. X. Weng, Y. Miao and G. Wang performed the simulations. Y. Wang, Q. Zhan and X. Dong analyzed all of the data. X. Weng and X. Gao co-wrote the paper, and J. Qu offered advice regarding its development. S. Zhuang directed the entire project. All authors discussed the results and contributed to the manuscript.

**Competing interests statement**

The authors declare that they have no competing financial or nonfinancial interests to disclose.

# Supplementary Information

## *Property unification of inherent amplitude, phase and polarization within a light beam*

Weng et al.

# Supplementary Note 1: Theoretical principle of Phase Malus's law

## Part 1: Physical ideas of Malus's law

In this note, we present the theoretical principle of Phase Malus's law in the main text. Firstly, we recall the physical idea of conventional Malus's law. Specially, when an incident linearly polarized light beam passes through a polarizer, the light intensity of output linearly polarized light beam possesses a one-to-one correspondence with the angle $\omega$ between the polarization direction of incident light beam and polarizer, as shown in Supplementary Fig. 1. Mathematically, this law implies the polarization-to-amplitude link in classical optics, which can be expressed as

$$I_o = | \mathbf{E}_0 |^2 \cos^2 \omega. \tag{1}$$

Therefore, according to the conservation of energy, the loss of light intensity can be obtained by $I_e = 1 - I_o = | \mathbf{E}_0 |^2 \sin^2 \omega$. Here, $\mathbf{E}_0$ is the amplitude of incident linearly polarized beam.

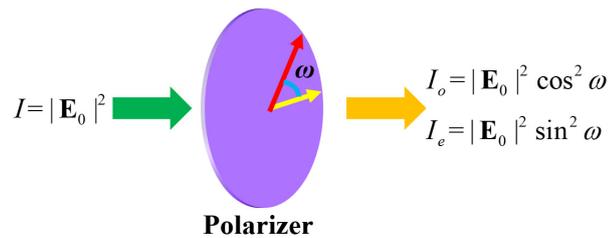

**Supplementary Figure 1.** Principle of conventional Malus's law.

From above, Malus's law implies two physical processes:

(1) **Energy transformation between $I_o$ and $I_e$.** Specifically, when the polarizer rotates, both parts of light intensities $I_o$ and $I_e$ transform each other by $\omega$. The increment of $I_o$ leads to the decrement of $I_e$. Therefore, there is an energy transformation between $I_o$ and $I_e$.

(2) **Scalar mode extraction of incident light beam.** $I_o$ and $I_e$ can be considered as two inherent scalar modes within the incident light beam. During propagating in free space, $I_o$ and $I_e$ are intertwined with each other, thereby maintaining the light beam stably. However, after passing through a polarizer, only the scalar mode with the same polarization direction of polarizer $I_o$ can be retained, while the other one $I_e$ is eliminated. From the point view of physics, Malus's law actually indicates an inherent scalar mode extraction of light beam.

## Part 2: Theoretical principle of Phase Malus's law

As discussed above, Malus's law implies scalar mode extraction from a single light beam. Therefore, its phase version, namely phase Malus's law, should also this mode extraction using the phase of light beam instead of a polarizer. Generally, the scalar modes $I_o = |\mathbf{E}_0|^2 \cos^2 \omega$ and $I_e = |\mathbf{E}_0|^2 \sin^2 \omega$ cannot be separated by modulating the phase of light beam in the wavefront, thereby a direct scalar mode extraction like that of using a polarizer cannot be realized in free space. Although a direct separation of scalar modes $I_o$ and $I_e$ cannot be achieved, an indirect solution can be realized using optical pen in the focal region of an objective lens[1].

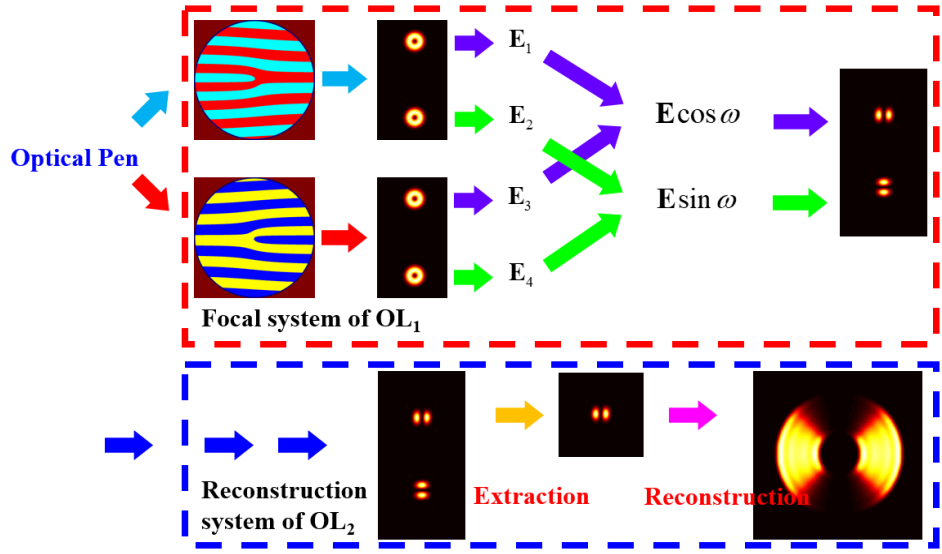

**Supplementary Figure 2.** Phase version of Malus's law. Phase version of Malus's law, namely phase Malus's law, is realized in Fig. 5, which can be composed with a focal system of OL$_1$ and a reconstruction system of OL$_2$. Scalar desired mode can be extracted by a pinhole in the focal region of OL$_1$, which is further reconstructed by OL$_2$. In this way, the phase B-to-amplitude is established.

**Focal system of OL$_1$**

Specifically, as shown in Fig. 5, when an incident light beam is focused by OL$_1$, the electrical fields near the focal point can be expressed as [2]

$$\mathbf{E}_f = \mathcal{F}\left\{P(\rho)Tl_0(\theta)\cos^{1/2}\theta \mathbf{M}\mathbf{E}_i \exp(-ik_z z)/\cos\theta\right\}, \tag{2}$$

where $\theta$ is the convergent angle, and the maximum convergent angle $\alpha = \arcsin(NA/n)$. $NA$ is the numerical aperture of OL$_1$, and $n$ is the refractive index in the focusing space. $k_z = k\cos\theta$ is the $z$ component of the wavevector $k = 2n\pi/\lambda$, where $\lambda$ is the wavelength of the incident light beam. $\mathcal{F}$ denotes the Fourier transform, and $P(\rho)$ represents the OL$_1$ aperture, which can be expressed as follows:

$$P(\rho) = \begin{cases} 1 & 0 < \rho < R \\ 0 & otherwise \end{cases}, \qquad (3)$$

where $R$ is the radius of OL₁. $T = \exp(i\phi)$, and the phase $\phi$ indicate the phase B. $l_0(\theta)$ denotes the electrical amplitude of the incident light beam. In this case, $l_0(\theta) = 1$.

In Supplementary Eq. (2), $\mathbf{E}_i$ represents the polarization state of incident light beam. For the sake of simplicity, the incident light beam is simplified into a superposition of the x and y polarized modes, expressed as follows:

$$\mathbf{E}_i = \cos\big[ mf(\theta, \varphi) \big]\langle \mathbf{x} \rangle + \sin\big[ mf(\theta, \varphi) \big]\langle \mathbf{y} \rangle, \qquad (4)$$

where $\langle \mathbf{x} \rangle$ and $\langle \mathbf{y} \rangle$ denote the x and y linearly polarized modes, respectively, and $\mathbf{M}$ denotes a polarization transformation matrix caused by OL₁, which can be expressed as [2]

$$\mathbf{M} = \begin{bmatrix} \cos^2\varphi\cos\theta + \sin^2\varphi & (\cos\theta-1)\sin\varphi\cos\varphi & -\sin\theta\cos\varphi \\ (\cos\theta-1)\sin\varphi\cos\varphi & \cos^2\varphi + \sin^2\varphi\cos\theta & -\sin\theta\sin\varphi \\ \sin\theta\cos\varphi & \sin\theta\sin\varphi & \cos\theta \end{bmatrix}. \qquad (5)$$

Eventually, the focal light intensity of the incident light beam can be obtained using $I = \left| \mathbf{E}_f \right|^2$.

**Two scaler modes using optical pen**

In the following simulations, $NA$=0.01, and $n$=1. The unit of length in all the figures is the wavelength λ, and the light intensity is normalized to the unit value. Because all light beams can be considered as the combination of the scalar mode $I_o = | \mathbf{E}_0 |^2 \cos^2\omega$ and $I_e = | \mathbf{E}_0 |^2 \sin^2\omega$, Phase B is valid for every light beam. Here, we take only a x linearly polarized beam as an example to create the scalar modes $I_o$ and $I_e$ in the focal region of OL₁. That is, $m = 0$ in supplementary Eq. (4).

In the above focal system, when modulating by the optical pen, the incident light beam can be divided into two complementary scaler modes in the focal region of OL₁ using optical pen [1]. Therefore, $T = \exp(i\phi_1)$ in supplementary Eq. (2), where $\phi_1$ can be expressed as [1]

$$\phi_1 = \text{Phase}\left( \sum\nolimits_{j=1}^{N} \text{PF}(s_j, f_j, \beta_j, z_j, \delta_j) \right) \qquad (6)$$

Here, $N$=2, $s_1 = s_2 = 1$; $\left( f_j, \beta_j, z_j \right)$ indicate the position of $j$-th focus with $-f_1 = f_2 = f$, $\beta_1 = \beta_2 = \eta$, $z_j = 0$; $\delta_j$ is the phase of $j$-th focus.

Mathematically, in case $\delta_1 = \omega$ and $\delta_2 = -\omega + 0.5\pi$, both complementary scaler modes can be expressed as

$$\mathbf{E}_1 = \mathbf{E}_0 \exp(i\omega) / 2 \tag{7}$$

$$\mathbf{E}_2 = i\mathbf{E}_0 \exp(-i\omega) / 2 \tag{8}$$

$\mathbf{E}_1$, $\mathbf{E}_2$ are the electrical amplitude of $j$-th focus. As shown in Supplementary Fig. 2, they are in the position of $(-f, \eta, 0)$ and $(f, \eta, 0)$, respectively.

In case $\delta_1 = -\omega$ and $\delta_2 = \omega + 0.5\pi$, both complementary scaler modes can be expressed as

$$\mathbf{E}_3 = \mathbf{E}_0 \exp(-i\omega) / 2 \tag{9}$$

$$\mathbf{E}_4 = -i\mathbf{E}_0 \exp(i\omega) / 2 \tag{10}$$

As shown in Supplementary Fig. 2, $\mathbf{E}_3$, $\mathbf{E}_4$ locate at the same position as that of $\mathbf{E}_1$, $\mathbf{E}_2$. However, their phases are adjusted to $-\omega$ and $\omega + 1.5\pi$, respectively.

From above, we can simply obtain the scalar modes $I_o = |E_0|^2 \cos^2 \omega$ and $I_e = |E_0|^2 \sin^2 \omega$ in the focal region of OL₁ by overlapping $\mathbf{E}_1$, $\mathbf{E}_2$ with $\mathbf{E}_3$, $\mathbf{E}_4$, respectively. Because the pinhole is placed in the focal plane of OL₁, we move $I_o$ to the geometric focus so that the phase-to-polarization link established by phase Malus's law is consistent with the phase-to-polarization link. Accordingly, the final phase of incident light beam, namely Phase B, can be expressed as

$$\phi_B = \text{Phase}\left(\left(Amp_a + Amp_b\right) \times \text{PF}(1, f, \eta, 0, 0)\right) \tag{11}$$

where

$$Amp_a = \text{PF}(1, -f, \eta, 0, \omega) + \text{PF}(1, f, \eta, 0, -\omega + 0.5\pi); \tag{12}$$

$$Amp_b = \text{PF}(1, -f, \eta, 0, -\omega) + \text{PF}(1, f, \eta, 0, \omega + 1.5\pi) . \tag{13}$$

In Supplementary Eqs. (12, 13), $\text{PF}(1, -f, \eta, 0, \omega)$, $\text{PF}(1, f, \eta, 0, -\omega + 0.5\pi)$ indicate the electrical amplitude $\mathbf{E}_1$, $\mathbf{E}_2$, respectively, while $\text{PF}(1, -f, \eta, 0, -\omega)$ and $\text{PF}(1, f, \eta, 0, \omega + 1.5\pi)$ denote the electrical amplitude $\mathbf{E}_3$, $\mathbf{E}_4$, respectively. Note that $I_o$ is moved to the geometric focus by $\text{PF}(1, f, \eta, 0, 0)$. Therefore, $I_e$ is located at $(2f, \eta, 0)$, and the entire optical system is compatible with polarized-SLM in Fig. 5.

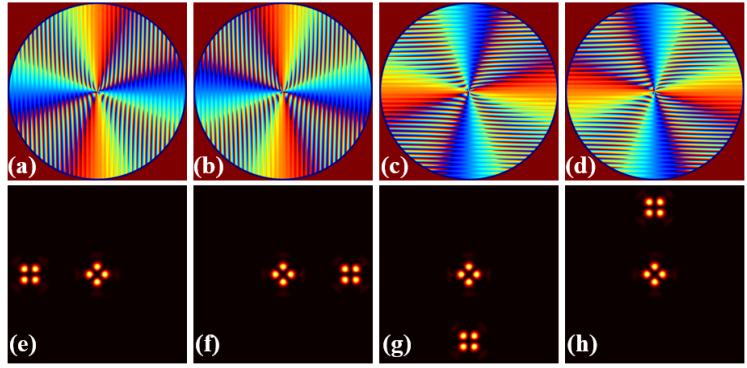

**Supplementary Figure 3.** The focal light intensity of $I_o$ and $I_e$ with different $\eta$ . Taking $\omega = 2\varphi$ as example. No matter what $\eta$ , the desired scalar mode $I_o$ remain the same in the focal region of OL$_1$.

Taking $\omega = 2\varphi$ as example in Supplementary Fig. 3. We present a series pairs of $I_o$ and $I_e$ with different $\eta$ . No matter what $\eta$ , one can always obtain $I_o = |E_0|^2 \cos^2 m\varphi$ and $I_e = |E_0|^2 \sin^2 m\varphi$ . Supposed that $I_o$ with different $\eta$ are overlapped together in the geometric focus, $I_o$ remains, while multiple $I_e$ are in the focal plane, as shown in Supplementary Fig. 3. Therefore, without loss of generality, we can extent the Phase B in Supplementary Eq. (11) into a generalized form, which can be expressed as

$$\phi_B = \text{Phase}\left(\sum_{j=1}^{N}\left(Amp_{aj} + Amp_{bj}\right) \times \text{PF}(1, f_j, \eta_j, 0, 0)\right) \tag{14}$$

where

$$Amp_{aj} = \text{PF}(1, -f_j, \eta_j, 0, \omega) + \text{PF}(1, f_j, \eta_j, 0, -\omega + 0.5\pi) \ ; \tag{15}$$

$$Amp_b = \text{PF}(1, -f_j, \eta_j, 0, -\omega) + \text{PF}(1, f_j, \eta_j, 0, \omega + 1.5\pi) \ . \tag{16}$$

Eventually, the Phase B in Supplementary Eq. (14) links with $I_o$ in the focal region of OL$_1$.

**Reconstruction system of OL$_2$**

In the focusing system of OL$_1$, we have demonstrated the spatial separation of $I_o = |E_0|^2 \cos^2 m\varphi$ and $I_e = |E_0|^2 \sin^2 m\varphi$ in the focal region using the phase in Supplementary Eq. (14) . Thus, one can simply extract the desired scalar mode $I_o$ by filtering out the undesired scalar mode $I_e$ using a pinhole. To establish the phase B-to-amplitude link, namely phase Malus's law, the desired scalar mode $I_o$ in the focal region of OL$_1$ is further reconstructed after passing through the objective lens OL$_2$, and its electric field can be expressed as [2]

$$\mathbf{E} = \left(P(\rho)\cos^{1/2}\theta\right)^{-1}\mathbf{M}^{-1}\exp(ik_z z)\cos\theta\mathcal{F}^{-1}\left(\mathbf{E}_p\right), \tag{17}$$

where $\mathcal{F}^{-1}$ denotes the inverse Fourier transform, and $\mathbf{E}_p$ is the electric field behind the pinhole in Supplementary Fig. 2, namely, the desired polarized mode. $\mathbf{M}^{-1}$ is the inverse polarization transformation matrix of the reconstructive lens OL₂, which can be expressed as [2]

$$\mathbf{M}^{-1} = \begin{bmatrix} \cos^2\varphi\cos\theta + \sin^2\varphi & (\cos\theta - 1)\sin\varphi\cos\varphi & \sin\theta\cos\varphi \\ (\cos\theta - 1)\sin\varphi\cos\varphi & \cos^2\varphi + \sin^2\varphi\cos\theta & \sin\theta\sin\varphi \\ -\sin\theta\cos\varphi & -\sin\theta\sin\varphi & \cos\theta \end{bmatrix}. \tag{18}$$

Finally, the light intensity output from OL₂ can be obtained using $I = |\mathbf{E}|^2$. Because the light intensity output from OL₂ is reconstructed from $I_o = |E_0|^2 \cos^2 m\varphi$ in the focal region, its also possesses an one-to-one correspondence with Phase B in Supplementary Eq. (14). In this way, the phase version of Malus's law is realized, thereby establishing the phase B-to-amplitude link in classical optics.

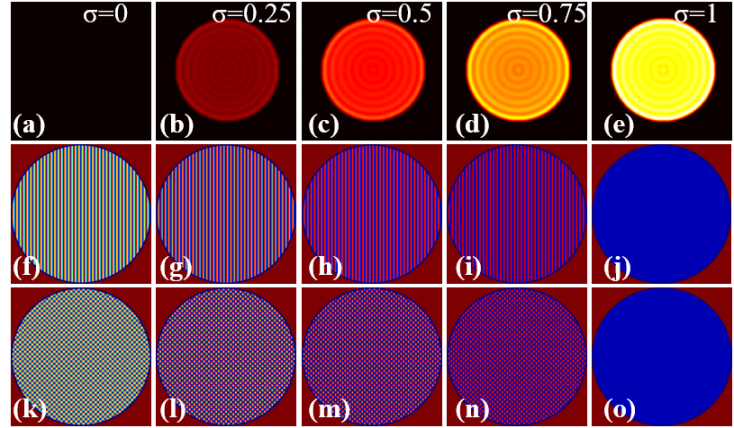

**Supplementary Figure 4.** Reconstruction light intensity from OL₂ with different amplitude $\sigma$. Here, subfigures (a-e) present light beams with amplitude (a) $\sigma = 0$; (b) $\sigma = 0.25$; (c) $\sigma = 0.5$; (d) $\sigma = 0.75$; (e) $\sigma = 1$, respectively. Their corresponding phase B are shown in (f-j) with *N*=1 and (k-o) with *N*=2, respectively.

To explain the principle of phase Malus's law more clearly, we calculate the reconstruction light intensity from OL₂ with different amplitude $\sigma$ in Supplementary Fig. 4. Here, we define the amplitude $\sigma = I_{out} / I_{in}$, where $I_{out}$ and $I_{in}$ are the light intensities of light beam output from OL₂ and incident light beam, respectively. Specifically, Supplementary Figures 4 (a-e) present light beams with (a) $\sigma = 0$; (b) $\sigma = 0.25$; (c) $\sigma = 0.5$; (d) $\sigma = 0.75$; (e) $\sigma = 1$, respectively. According to Supplementary Eq. (14), one identical $\sigma$ can link to different phases, which is determined by the parameters $N$ and $f_j$, $\eta_j$. Therefore, different amplitudes $\sigma$ in Supplementary Figs. 4 (a-e) are corresponding to multiple phases in Supplementary Figs. 4 (f-j) with *N*=1 and (k-o) with *N*=2, respectively. Their only difference is the different position of undesired scalar modes $I_e$ in the focal region of OL1, see Supplementary Fig. 3. However, it does not affect the inner desired scalar mode. For this reason, although their phase forms are different from each other, they still can be classified as a kind of phase, namely Phase B. Based on the above discussion, the essence of phase Malus's law is to transfer the amplitude change into a kind of phase in Supplementary Eq. (14), as shown in Fig. 3.

**Extension of Phase Malus's law**

Phase Malus's law demonstrates a fundamental link between Phase B in Supplementary Eq. (14) and $I_o = |\mathbf{E}_0|^2 \cos^2 \omega$. For $\omega = \omega_0$ is a const, the amplitude of entire light beam can be adjusted to a constant $\sigma$; For $\omega$ is a variate, one can obtain a localized amplitude modulation of light beam in free space by $I_o = |\mathbf{E}_0|^2 \cos^2 \omega$. The first case indicates a similar function like that of amplitude modulation using a polarizer. The second case implies a pixelate modulation based on the principle of Phase Malus's law. However, according to the phase B-to-polarization link, it is difficult to achieve both functions simultaneously in free space. Therefore, we extent Phase Malus's law to a more generalized form so that one can achieve the relationship between the phase B and $I_o = |\mathbf{E}_0|^2 A^2 \cos^2 \omega$. Here, $A = \cos \omega_0$ and the range of $\omega_0$ is 0~0.5π.

Without loss of generality, we only take $N=1$ in Supplementary Eq. (14) as example to explain the additional function of phase Malus's law. Mathematically, the scalar mode $I_o = |\mathbf{E}_0|^2 A^2 \cos^2 \omega$ can simply obtained by the phase $\phi = \phi_{B1} + \phi_{B2}$, where

$$\phi_{B1} = \text{Phase}\left(\left(Amp_{a1} + Amp_{b1}\right) \times \text{PF}(1, f_1, \eta_1, 0, 0)\right) \tag{19}$$

$$\phi_{B2} = \text{Phase}\left(\left(Amp_{a2} + Amp_{b2}\right) \times \text{PF}(1, f_2, \eta_2, 0, 0)\right) \tag{20}$$

$Amp_{a1}$, $Amp_{b1}$ can be found in Supplementary Eqs. (15, 16). $\eta_1 \neq \eta_2$. For the former phase, $\phi_{B1}$ are corresponding to $I_1 = |\mathbf{E}_0|^2 \cos^2 \omega_0$, thereby leading to a constant amplitude modulation of entire output light beam. For the latter phase, $\phi_{B2}$ links with $I_2 = |\mathbf{E}_0|^2 \cos^2 \omega$ directly, where pixelate modulation of light beam can be realized by the variate $\omega$. Therefore, one can simply obtain $I_o = |\mathbf{E}_0|^2 A^2 \cos^2 \omega$ by the overall phase $\phi$ of incident light beam. Here, $A = \cos \omega_0$.

Supplementary Figure 5 presents two theoretical results of $I_o = |E_0|^2 A^2 \cos^2 \omega$. One is a folk pattern generated by the phases in Supplementary Figures 5 (a-e), where their corresponding amplitudes

$\sigma$ are (f) $\sigma = 0$; (g) $\sigma = 0.25$; (h) $\sigma = 0.5$; (i) $\sigma = 0.75$; (j) $\sigma = 1$, respectively; another is a spiral pattern with (q) $\sigma = 0$; (r) $\sigma = 0.25$; (s) $\sigma = 0.5$; (t) $\sigma = 0.75$; (u) $\sigma = 1$, which are created by the phases in Supplementary Figures 5 (k-o), respectively. According to these theoretical results, one can found that arbitrary light intensity distribution can be obtained simply by adjusting the phase $\phi_{B2}$, while the phase $\phi_{B1}$ acts as a polarizer that can manipulate the entire light intensity on demand. This addition function offers a new alternative for adjusting the light intensity of arbitrary light pattern in a digital way.

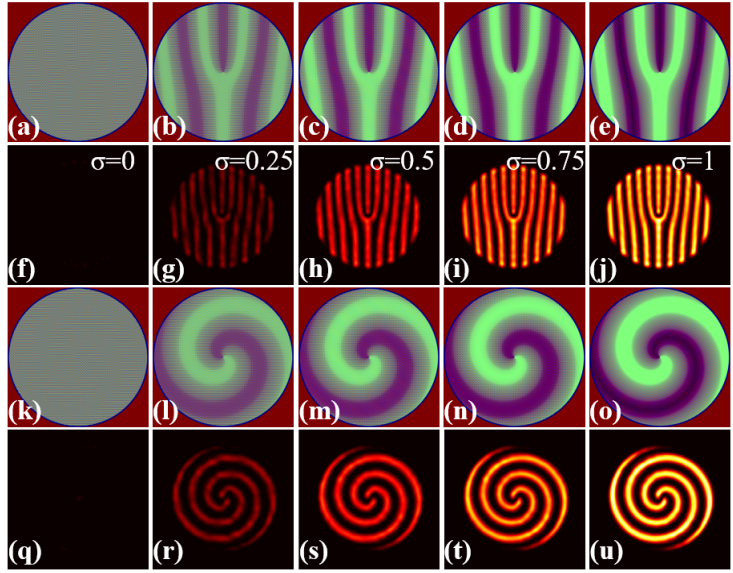

**Supplementary Figure 5.** Theoretical results of $I_o = |\mathbf{E}_0|^2 A^2 \cos^2 \omega$. Here, (f-j) and (q-u) are folk patterns and spiral patterns with different $\sigma$, where (f, q) $\sigma = 0$; (g, r) $\sigma = 0.25$; (h, s) $\sigma = 0.5$; (i, t) $\sigma = 0.75$; (j, u) $\sigma = 1$, respectively. Their corresponding phases are shown in (a-e) and (k-o), respectively.

## Supplementary Note 2: Theoretical principle Full-property SLM

Based on the unification of amplitude, phase and polarization in Fig. 4, Full-property SLM can be simplified to a filter system in Fig. 5. Therefore, the theoretical principle of Full-property SLM is similar to that of Supplementary Note 1. Their differences are merely the different polarization state and phase of incident light beam.

Here, the incident light beam is a $m$=30 order VVB. Therefore, in Supplementary Eq. (2), its polarization state $\mathbf{E}_i$ can be expressed as

$$\mathbf{E}_i = \cos m\varphi \langle \mathbf{x} \rangle + \sin m\varphi \langle \mathbf{y} \rangle, \tag{21}$$

where $\langle \mathbf{x} \rangle$ and $\langle \mathbf{y} \rangle$ denote the x and y linearly polarized modes, respectively. Due to the slight deviation between the conjugated plane of the phase-only SLM and VP, the output vector beam from the Full-property SLM has a hollow shape, as shown in Fig. 6. In this case, the electrical amplitude $l_0(\theta)$ of the incident VVB can be approximately expressed as follows:

$$l_0(\theta) = \begin{cases} 0 & 0 < \sin\theta / NA < 0.2 \\ 1 & 0.2 \leq \sin\theta / NA < 1 \end{cases}. \tag{22}$$

$T = \exp(i\phi)$ and the phase $\phi$ indicate the phase of incident $m$ order VVB, which is composed with three different kinds of phase in Eq. (7), namely Phase A, B and C, respectively. As discussed in the main text, Phase A, B and C link to the polarization, amplitude and phase of $m$ order VVB, thereby making it possible to modulate these three properties merely by its scalar phase.

By substituting Eq. (7) and Supplementary Eq. (21), (22) into Supplementary Eq. (17), one can simply obtain all the theoretical results in Fig. 6. Here, some common parameter of Phase B in Eq. (3) are $N$=1; $s_j$=1; $\beta_{j,} = 0.25\pi$; $f_j = sk / NA$ with $sk$=30, $NA$=0.01; $k$=2$\pi$.

**Supplementary References**